\documentclass[12pt,preprint]{aastex}
\shorttitle{Bullet Cluster in cDE Models}
\shortauthors{Lee \& Baldi}
\begin{document}
\title{Dark Sector Coupling Bends the Superclusters}
\author{Junsup Shim and Jounghun Lee}
\affil{Astronomy program, Department of Physics and Astronomy,
Seoul National University, Seoul 151-747, Republic of Korea \\
\email{jsshim@astro.snu.ac.kr, jounghun@astro.snu.ac.kr}}
\begin{abstract}
The galaxy clusters exhibit  noticeably anisotropic pattern in their clustering, which is vividly manifested 
by the presence of rich filament-like superclusters. The more anisotropic the clustering of galaxy clusters 
is, the more straight the rich filament-like superclusters become. Given that the degree of the anisotropy 
in the largest-scale clustering depends sensitively on the nature of dark energy, the supercluster straightness 
may play a complimentary role in testing dynamic dark energy models. Here we focus on the coupled dark 
energy (cDE) models which assume the existence of dark sector coupling between scalar field dark energy 
and nonbaryonic dark matter. Determining the spines of the superclusters identified in the publicly available  
group catalogs from the CODECS (COupled Dark Energy Cosmological Simulations) for four different 
cDE models as well as for the $\Lambda$CDM model, we quantify the straightness of each supercluster 
as the spatial extent of its spine per member 
cluster where a supercluster spine represents the main stem of the minimal spanning tree constructed 
out of the member clusters. It is shown that the dark sector coupling plays a role in making the supercluster less 
straight relative to the $\Lambda$CDM case and that  in a cDE model with supergravity potential the 
superclusters are least straight. We also find that the difference in the degree of the supercluster straightness 
between the cDE and the $\Lambda$CDM cases increases with redshifts. A physical interpretation of our result 
as well as its cosmological implication are discussed. 
\end{abstract}
\section{INTRODUCTION}

The meticulous analysis of the CMB (Cosmic Microwave Background) temperature power spectrum done by the 
Planck mission team has casted both light and shadow on the status of  the $\Lambda$CDM (cosmological 
constant $\Lambda$ + cold dark matter) cosmology.  While at high multipoles ($l\ge 500)$ the $\Lambda$CDM 
cosmology is found perfect in matching the observational data,  at low multipoles ($l\le 100$) its poor-fits has 
been confirmed not as numerical flukes but likely to be real \citep{planck-para13,planck-cmb13}. Given that the 
low-multipole behavior of the CMB temperature spectrum reflects the most primordial feature of the universe,  
the Planck result along with the infamous long-standing problem associated with $\Lambda$  may imply the 
incompleteness of the $\Lambda$CDM cosmology,  encouraging the cosmologists to search more strenuously 
than ever for physical alternatives.

The coupled dark energy (cDE) models where dark energy is not the inert $\Lambda$ but a dynamic scalar field 
coupled to  nonbaryonic dark matter particles \citep{wetterich95,amendola00,amendola04,PB08,baldi-etal10} 
have recently attained probing attention because of their capacity for accommodating several observational 
mysteries that the $\Lambda$CDM cosmology could not resolve. For instance, according to \citet{baldi12a}, the 
high-$z$ massive clusters regarded as extremely rare events in the $\Lambda$CDM universe \citep[e.g., see][and 
references therein]{jee-etal11} are more probable to detect in the cDE models.   \citet{BLM11} demonstrated that 
the cDE models can explain the observed higher degree of the misalignment between the spatial distributions of 
cluster galaxies and  dark matter than naturally expected in the $\Lambda$CDM cosmology
\citep{oguri-etal10,lee10}.  In the work of \citet{LB12},  the morphological properties of the observed bullet 
cluster \citep{clowe-etal04,clowe-etal06,markevitch-etal02,markevitch-etal04,markevitch-etal05,MB08} that had 
been found to be in a serious tension with the prediction of the $\Lambda$CDM cosmology 
\citep{FR07,LK10, TN12,AY12}  were much less anomalous in cDE models.  
Very recently, \citet{SM13} claimed that the dispute on the value of the Hubble constant between the Planck 
experiment and the HST (Hubble Space Telescope) project can be ended if cDE is assumed to exist. This 
capacity of the cDE models to alleviate the tensions posed by the observational anomalies is mainly owing to the 
presence of an additional  long-range fifth force induced by the dark sector coupling (the interaction between 
dark energy and nonbaryonic dark matter). See \citet{AT10} for a comprehensive review of the cDE models.

Since the tight constraints put on the strength of dark sector coupling by the recent observations have sustained 
as viable only those cDE models which are very hard to distinguish from the $\Lambda$CDM cosmology 
\citep[e.g.,][]{bean-etal08},  it is necessary to develop as a powerful indicator of cDE as possible for the 
detection of  the effect of cDE on the evolution of the universe.  The spatial clustering of galaxy clusters that 
generates collectively the cosmic web phenomenon \citep{web96} has been regarded as one of the most 
powerful indicators of dynamic dark energy.  As the dark sector coupling would affect not only the strength of 
the large-scale clustering but also its degree of anisotropy, the cosmic web must take on different pattern in the 
presence of cDE.  

The anisotropic clustering of clusters is well manifested by the elongated filamentary shapes of the rich 
superclusters which correspond to the densest section of the cosmic web. 
In the literatures which studied the superclusters and their morphological properties,  the shapes of the 
superclusters were measured by various different algorithms such as percolation, ellipsoid-fitting, 
friends-of-friends, Minkowski functional and etc. 
\citep[e.g.,][]{dekel-etal84,west89,plionis-etal92,jaaniste-etal98,basilakos-etal01,basilakos03,
einasto-etal07,wray-etal06,einasto-etal11}.
The general consensus of those previous works was that no matter what algorithm was used, the richer 
superclusters appear to have more filamentary shapes.  Recently, \citet{einasto-etal11} noted that the shape of 
a richest supercluster located in the highly overdense region is best described as   a "multi-branch" filament 
consisting of the main stem and several branches. 

Assuming that the degree of the straightness of the supercluster main stems would depend strongly on the 
dynamics of dark energy, we speculate that in cDE models the main stems of rich superclusters would be less 
straight, having shorter spatial extents compared with the $\Lambda$CDM case, due to the effect of the fifth 
force. To quantitatively inspect this speculation, we utilize the data from the high-resolution N-body simulations 
ran for various cDE models as well as for the $\Lambda$CDM model.  
The upcoming chapters are organized as follows: In section 2, we briefly review the cDE scenarios and describe 
how the spatial extents of the superclusters found in the N-body data are determined.  
In section 3, we show how the degree of the supercluster straightness depends on the strength of dark sector 
coupling and how the difference in the supercluster straightness between the cDE and the $\Lambda$CDM 
cases changes with redshifts.  In section 5, we summarize the key results and draw a final conclusion.

\section{NUMERICAL DATA  AND ANALYSIS}

\subsection{A Brief Summary of the CODECS}

The CODECS stands for the COupled Dark Energy Cosmological Simulations performed by \citet{baldi12b} 
for several different cDE models as well as for the standard $\Lambda$CDM model with $1024^{3}$ CDM 
particles and the same number of baryon particles in a periodic box of linear size $1\,h^{-1}$Gpc.  The 
CODECS has the mass-resolution of $m_{\rm c}=5.84\times 10^{10}\,h^{-1}M_{\odot}$ and 
$m_{\rm b}=1.17\times 10^{10}\,h^{-1}M_{\odot}$ where $m_{\rm c}$ and $m_{\rm b}$ represent the mass of 
each CDM and baryon particle at the present epoch, respectively. As conventionally done, a bound dark halo 
was identified in the CODECS suite as a  friends-of-friends (FoF) group in which the component particles are 
all within the linking length of $0.2\bar{l}$ where $\bar{l}$ is the mean particle separation \citep{FoF02}. In the 
CODECS project the initial conditions of all models were set to  be consistent with the WMAP7 values 
\citep{wmap7}. For the cDE models the normalization amplitude of the density power spectrum, $\sigma_{8}$, 
has the same value as the $\Lambda$CDM case at the moment of the last-scattering, while  the 
other key parameters have the WMAP7 values at $z=0$ .  For a full description of the CODECS, see 
\citet{baldi12b}.

Among several target cDE models of the CODECS,  the following four models are considered for our analysis: 
EXP002, EXP003, EXP008e3 and SUGRA003. In the first two models (EXP002 and EXP003) where the 
dynamics of a scalar field cDE, $\phi$, is governed by an exponential potential of 
$U(\phi)\propto \exp(-\alpha\phi)$ with $\alpha=0.08$ \citep{LM85,RP88,wetterich88}, the coupling parameter 
$\beta$ that quantifies the strength of the dark sector coupling has a positive constant value of $0.1$ and $0.15$, 
respectively. In the third model (EXP008e3) where the cDE potential has the same exponential shape, the 
coupling strength is no longer a constant but depends exponentially on $\phi$ as $\beta=0.4\exp(3\phi)$ . In the 
fourth model (SUGRA003) where the cDE evolves according to the supergravity potential of 
$U(\phi)\propto\phi^{-\alpha}\exp(\phi^{2}/2)$ \citep{BM99}, $\beta$ has a negative constant value of $-0.15$ 
\citep{baldi12a}. The amplitudes of the linear density power spectrum at $z=0$ have the values of 
$\sigma_{8}=0.875,\ 0.967,\ 0.895$ and $0.806$ at $z=0$ for the EXP002, EXP003, EXP008e3 and SUGRA003 
models, respectively. For comparison, we also consider the $\Lambda$CDM model included in the CODECS 
project for which $\sigma_{8}=0.809$. For a detailed explanation of the cDE models considered in the 
CODECS, see \citet{baldi12b}.

\subsection{Identifying the Supercluster Spines}

\begin{figure}[ht]
\begin{center}
\plotone{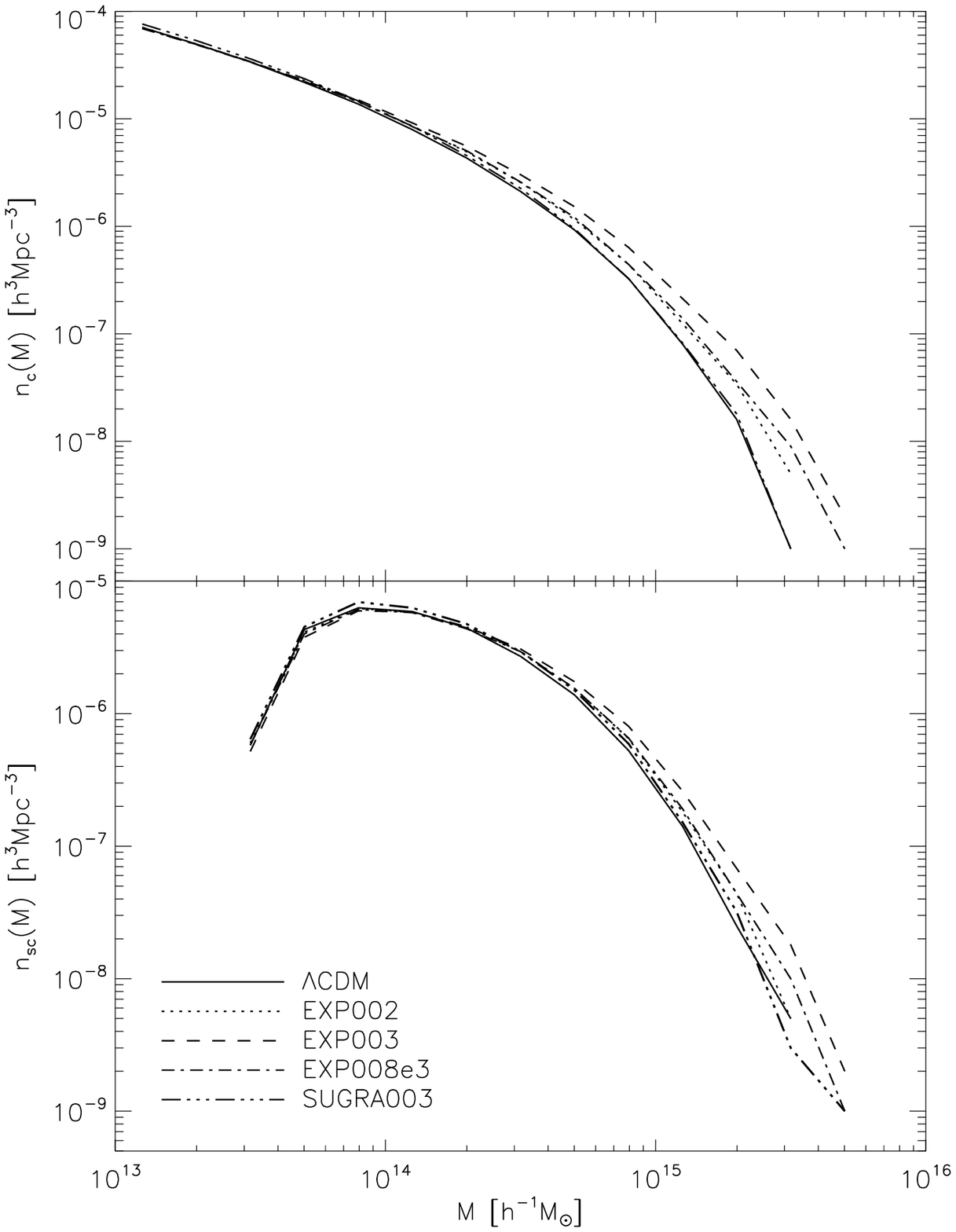}
\caption{Number densities of the selected cluster halos (top panel) and the 
superclusters identified as the FoF groups of the selected cluster halos with linkage parameter of $1/3$ 
(bottom panel)  per unit volume at $z=0$ for five different cosmological models from the CODECS. }
\label{fig:mf}
\end{center}
\end{figure}

From the dark halo catalogs from the CODECS at $z=0$,  we first  select only those cluster halos whose 
masses $M$ satisfy the condition of $M\ge 10^{13}\,h^{-1}M_{\odot}$ for each model.  Following the 
conventional scheme \citep[e.g.,][]{KE05,lee06,wray-etal06,LK10}, we identify the superclusters as the 
FoF groups of the closely located clusters among the selected ones within the linking length, $l$, equal to one 
third of the mean separation of the selected clusters, $\bar{d}_{\rm c}$,  as $l=\bar{d}_{\rm c}/3$. Figure 
\ref{fig:mf} plots the number densities of the selected clusters and the identified superclusters per unit volume for 
five cosmological models at $z=0$ in the top and bottom panels, respectively. As can be seen, the selected 
clusters of each model exhibit a different mass spectrum. This result indicates that the selected clusters of each 
model has a different mean separation $\bar{d}_{\rm c}$, which has been in fact properly taken into account 
when the superclusters are identified as FoF groups of the linked clusters within the distance of $l$. Note that the 
massive clusters and superclusters are most abundant in the EXP003 model, while the SUGRA003 model 
displays almost the same mass distributions as the $\Lambda$CDM case.

Choosing only those rich superclusters which have three or more member clusters,  we apply the minimal 
spanning tree (MST) technique to dig out their filamentary patterns for each cosmological model.  It was 
\citet{barrow-etal85} who for the first time used the MST algorithm to study systematically the local galaxy 
distribution.  Afterwards,  the  MST algorithm has been refined and modified by several authors to investigate 
the geometrical properties of the large scale structure with it 
\citep[e.g.,][]{KS96,doroshkevich-eta01,colberg07,PL09}.  

It may be worth explaining here why we choose the MST algorithm for our analysis rather than more recently 
developed algorithms such as the Skeleton \citep{skeleton08}, the Multiscale Morphology Filter \citep{mmf10}, 
the NEXUS \citep{nexus13},  and so forth, all of which are believed to be more accurate and efficient in tracing 
the cosmic web. The number one reason is that the application of those more recent and more elaborate 
algorithms require us to know the underlying dark matter distribution (or density/velocity fields) unlike the MST 
algorithm for which only the halo distributions are necessary. Given that the dark matter distribution from the 
CODECS is not publicly available, we had to find an algorithm based only on the halo distributions. 
Of course, it is still possible in principle to reconstruct the density/velocity fields from the CODECS halo 
catalogs and then to apply those more improved algorithms to the reconstructed density/velocity fields. 
But,  our ultimate goal is not to model as accurately as possible the supercluster shapes but to see if and how 
the supercluster straightness depends on the background cosmology. Thus, we belive that the 
reconstruction of the density fields of all five models is beyond the scope of this paper. In fact, the MST which 
treats each member cluster as a point (node) without weighing it by its mass must be the most optimal (and the 
most practical) algorithm to achieve our goal since we would like to separate the cosmology dependence of the 
supercluster straightness from that of the supercluster mass which is anyway hard to estimate accurately in real 
observation.

\begin{figure}[ht]
\begin{center}
\plotone{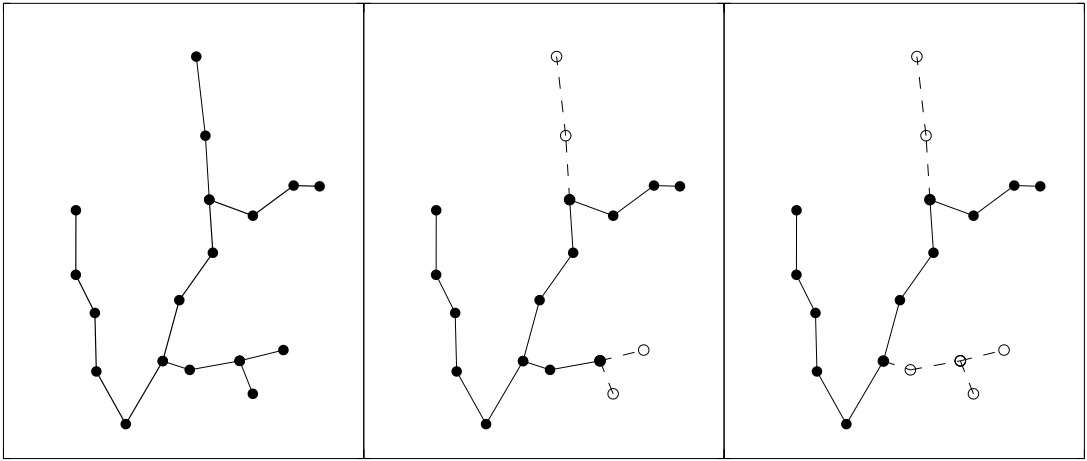}
\caption{Illustration of the pruning process of a supercluster with $18$ nodes identified in the CODECS group 
catalog at $z=0$ in the two dimensional plane. All the branches composed of two or less nodes are regarded as 
not a part of the main stem and thus  cut down by the pruning process. }
\label{fig:prune}
\end{center}
\end{figure}

\begin{figure}
\begin{center}
\plotone{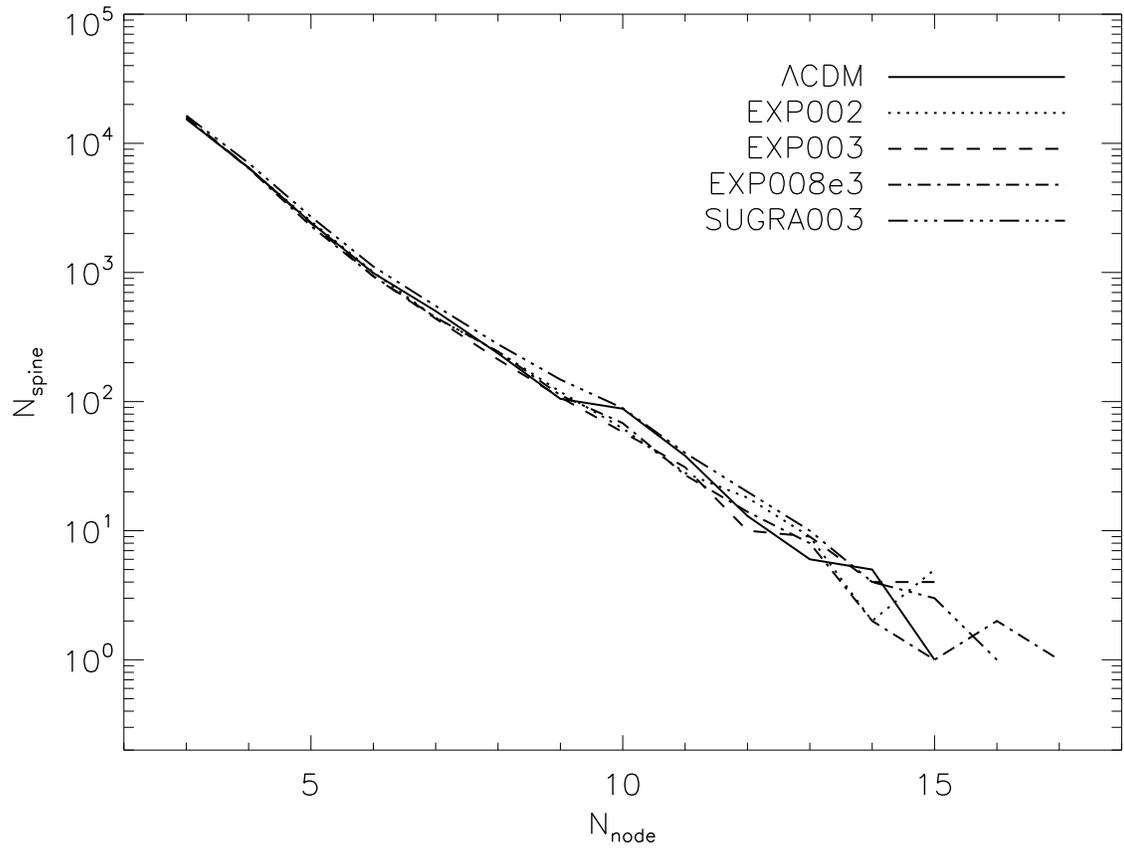}
\caption{Node number distributions of the supercluster spines for five different models at $z=0$.}
\label{fig:pnode}
\end{center}
\end{figure}

After constructing the MST of each supercluster,  we extract its main stem by pruning off its minor branches, 
under the assumption that the main stem of a supercluster MST corresponds to its most prominent 
filamentary part. Since we confine the MST reconstruction procedure to the member cluster distribution inside 
each supercluster, we only prune a supercluster MST without separating it into smaller filamentary parts 
\citep[c.f.,][]{barrow-etal85,colberg07,PL09}.  Figure \ref{fig:prune} illustrates in the two-dimensional projected 
space how the main stem of a supercluster MST is determined through pruning process at $z=0$ for the 
$\Lambda$CDM model, displaying how the minor branches are repeatedly cleared off from 
the main stem of a supercluster MST.  As \citet{colberg07} called the main stem of a MST out of the galaxy 
distribution as the "backbone of the large-scale structure", hereafter we also call the main stem of a supercluster 
{\it the supercluster spine}. Figure \ref{fig:pnode} plots the number distribution of the member clusters of 
those rich supercluster spines consisting of three or more nodes for the five models at $z=0$.  As can be seen, 
the distributions of the node numbers ($N_{\rm node})$ of the supercluster spines for the five models are very 
similar to one another except for the numerical fluctuations in the large-$N_{\rm node}$ section, even though 
each model has a different mass distribution (see Figure \ref{fig:mf}). 

\begin{figure}[ht]
\begin{center}
\plotone{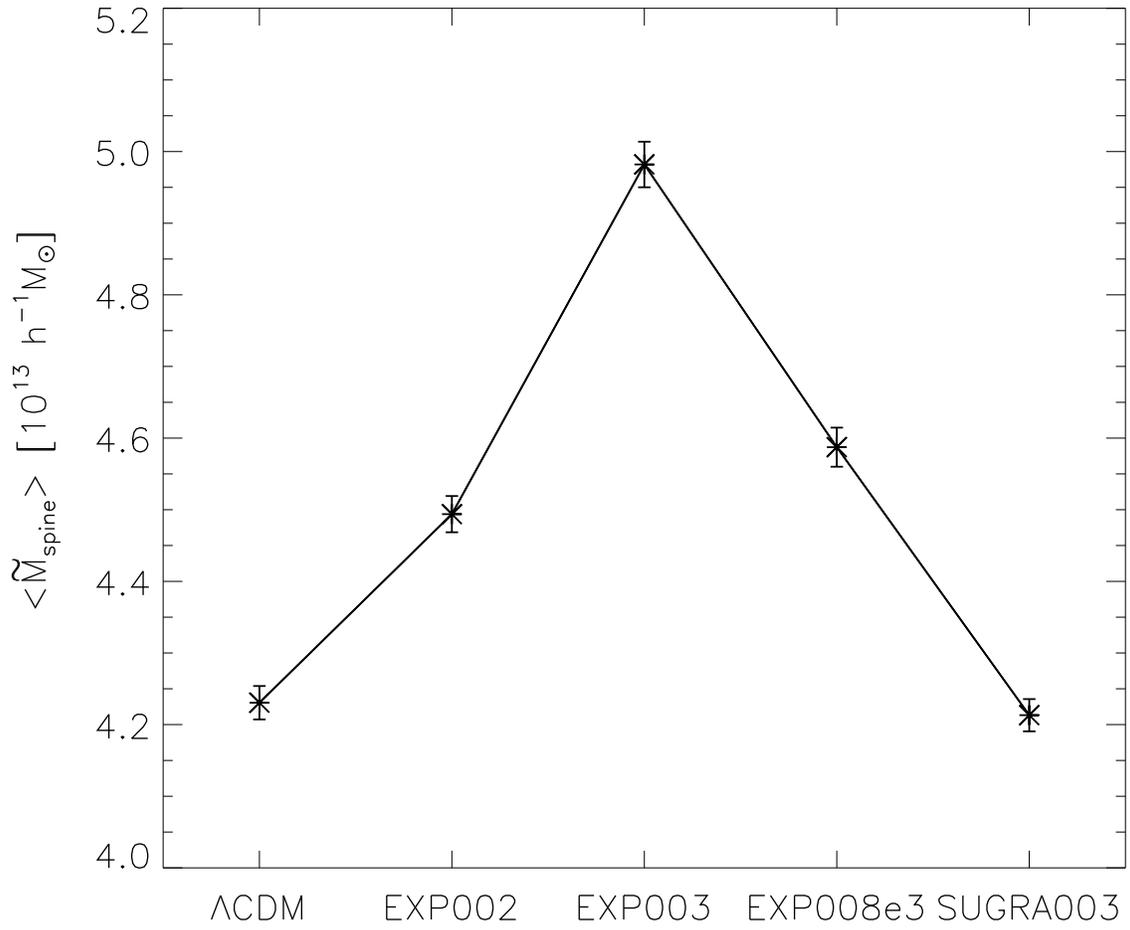}
\caption{Mean specific masses of the supercluster spines at $z=0$ vs. the cosmological models. }
\label{fig:spinesm}
\end{center}
\end{figure}

An acute reader might concern about a possibility that in case of a more or less spherically shaped supercluster 
the pruning process would remove its most massive member from the main stem. We have investigated how 
probable this case is for each model and found that the fraction of all supercluster MSTs occupied by those 
cases is less $0.03$ for every model, having very negligible effect on the final result.  Table \ref{tab:mmass} lists 
the numbers of those rich superclusters whose spines consist of three or more nodes and the percentage of the 
supercluster spines from which the most massive member clusters are pruned away at $z=0$ for the five 
models. Figure \ref{fig:spinesm} plots the mean specific mass, 
$\langle \tilde{M}_{\rm spine}\rangle\equiv M_{\rm spine}/N_{\rm node}$,  of the rich supercluster spines vs. the 
cosmological model. As can be seen, the mean specific mass is the highest for the EXP003 case, indicating that 
the superclusters tend to have higher masses when the amplitude of the density power spectrum, $\sigma_{8}$, 
has a higher value. 

\section{SUPERCLUSTER STRAIGHTNESS AS A PROBE OF cDE}

\begin{figure}
\begin{center}
\plotone{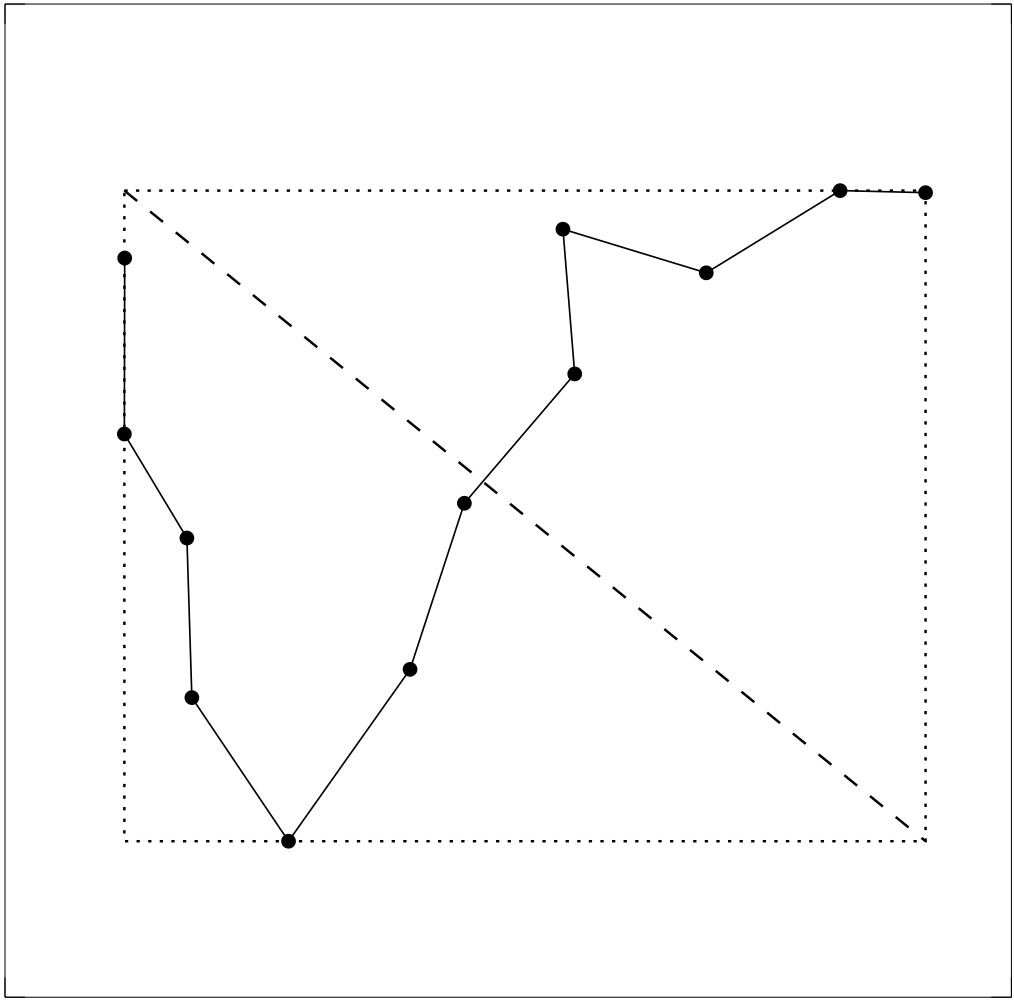}
\caption{Size (dashed line) of a supercluster spine in the two dimensional projected plane.}
\label{fig:size_mst}
\end{center}
\end{figure}

Using only those rich superclusters whose spines consist of three or more nodes, we 
determine their sizes, $S$.  Although a detailed explanation about how to measure the size of a pruned MST is 
provided in \citet{PL09} \citep[see also,][]{colberg07}, we also briefly describe here the procedure to 
estimate the size of each supercluster spine to make this paper self-contained.  Let the comoving 
Cartesian coordinates of all nodes belonging to a supercluster spine be in range of 
$x_{\rm min}\le x\le x_{\rm max},\ y_{\rm min}\le y\le y_{\rm max},\ 
z_{\rm min}\le z\le z_{\rm max}$, respectively.  The size, $S$, of a supercluster spine is now estimated as 
$S=\left[(x_{\rm max}-x_{\rm min})^{2}+(y_{\rm max}-y_{\rm min})^{2}+
(z_{\rm max}-z_{\rm min})^{2}\right]^{1/2}$, which quantifies effectively how extended the supercluster spine 
is in the three dimensional space.  Figure \ref{fig:size_mst} illustrates how the size $S$ of a supercluster spine is 
measured at $z=0$ for the $\Lambda$CDM case in the two-dimensional projected space.  

\begin{figure}
\begin{center}
\plotone{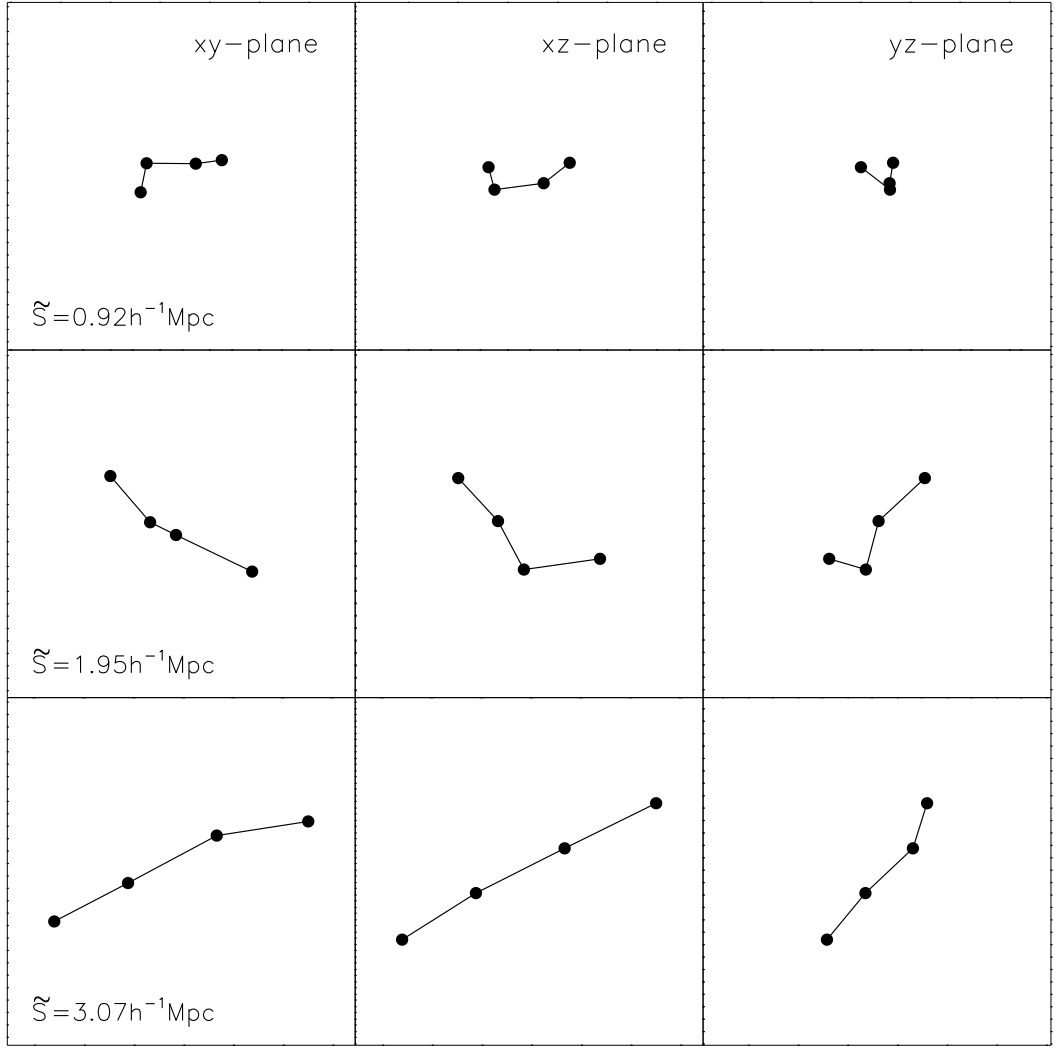}
\caption{Projected images of three randomly chosen supercluster spines having same number of nodes but  
different specific sizes from the $\Lambda$CDM case at $z=0$.}
\label{fig:straight}
\end{center}
\end{figure}

Obviously the size of a supercluster spine increases with the number of nodes. When the number of nodes 
$N_{\rm node}$ is fixed, however, the more straight superclusters should have larger sizes.
To quantify its degree of the straightness of a supercluster spine, we define the specific size $\tilde{S}$  as the 
size per node, $\tilde{S}\equiv S/N_{\rm node}$. Figure \ref{fig:straight} shows the projected images of three 
randomly chosen supercluster spines with three different specific sizes at $z=0$ for the $\Lambda$CDM 
case. As can be seen, those supercluster spines with larger specific sizes look more straight, demonstrating 
that the specific size of a supercluster spine is an effective indicator of its straightness.

\begin{figure}[ht]
\begin{center}
\plotone{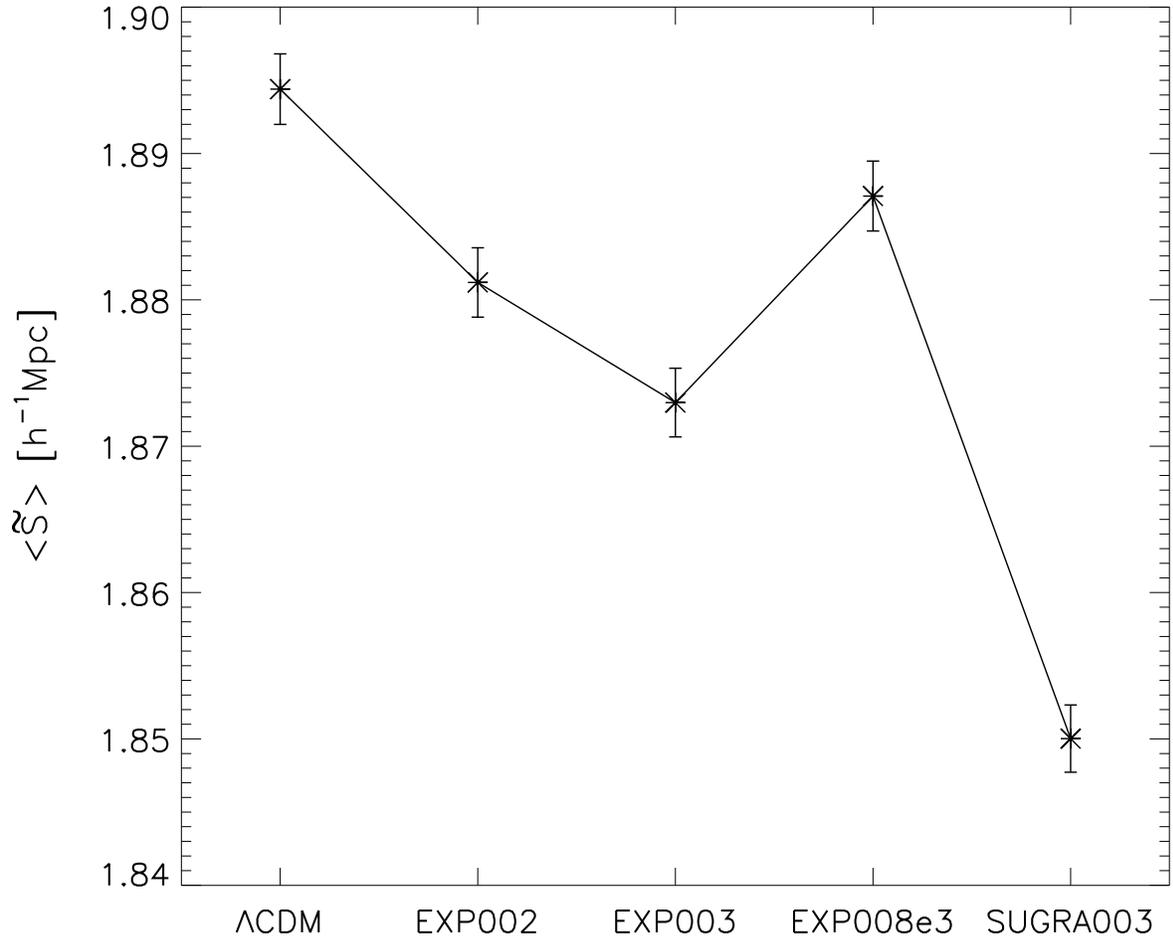}
\caption{Mean specific sizes of the supercluster spines for the five models at $z=0$. }
\label{fig:mssize}
\end{center}
\end{figure}

Now, we  estimate the mean value of $\tilde{S}$ averaged over those superclusters whose spines have 
$N_{\rm node}\ge 3$ for each cosmological model. Figure \ref{fig:mssize} plots the mean specific sizes, 
$\langle\tilde{S}\rangle$, versus the five cosmological models at $z=0$. The errors $\sigma_{\tilde{S}}$ are 
calculated as the one standard deviation in the measurement of the mean value 
$\sigma_{\tilde{S}}=[(\langle\tilde{S}^{2}\rangle-\langle\tilde{S}\rangle^{2})/(N_{\rm spine}-1)]^{1/2}$.
As can be seen, in the cDE models the supercluster spines tend to have smaller specific sizes. In other words, 
the superclusters in models with cDE are less straight. 
The $\Lambda$CDM case exhibits the highest value of $\tilde{S}$ while the lowest value is found for the 
SUGRA003 case.  As can be seen, the difference in $\tilde{S}$ between the $\Lambda$CDM and the 
SUGRA003 cases is the most significant.  There is a clear trend that the mean value of $\tilde{S}$ 
decreases with increasing coupling strength for the cases that cDE has a constant constant coupling (EXP002 
and EXP003) while the EXP008e3 case does not show a significant difference in $\tilde{S}$ from the 
$\Lambda$CDM case. 

It is important to note that the cosmology dependence of $\langle\tilde{S}\rangle$  is obviously different from that 
of $\langle \tilde{M}_{\rm spine}\rangle$ by comparing the result shown in Figure \ref{fig:spinesm} with that in 
Figure \ref{fig:mssize}.   The mean specific mass of the supercluster spines in the SUGRA003 model turns out 
to be very similar to the $\Lambda$CDM case, while the two models show significant difference in the 
mean specific size of the supercluster spines.  This result implies that the difference in the supercluster mass 
existent among the models should not be the cause of the detected overall trend between the cDE models and 
the mean specific sizes of the supercluster spines. 

The result shown in Figure \ref{fig:mssize} reveals that the dark sector coupling plays a role in diminishing the 
degree of the straightness of the superclusters. Our interpretation is as follows: The accelerating expansion 
of the Universe caused by the anti-gravitational action of dark energy sharpens the cosmic web while the 
gravitational clustering of clusters tends to blunt it since the former (latter) increases the relative dominance of 
the anisotropic (isotropic) stress on the supercluster scales. The competition between the two driving forces 
determines the degree of the straightness of the superclusters which reflects how sharp the cosmic web is 
in the universe. In the cDE models, the additional attractive fifth force with long range helps gravity 
blunt the cosmic web on the large scale, reducing the specific sizes of the supercluster spines, which 
is why the stronger dark sector coupling makes the supercluster spines less straight.

The lowest value of $\langle\tilde{S}\rangle$ found for the SUGRA003 case may be also interpreted as follows. 
The degree of the supercluster straightness is also affected by the peculiar velocity perturbation.  The large 
peculiar velocity perturbation tends to sharpen the cosmic web. A good analogy can be found in a warm dark 
matter (WDM) model where the peculiar velocity perturbation of DM particles is much larger than 
for the CDM case. \citet{GT07} clearly demonstrated by using a high-resolution hydrodynamic 
simulation that a WDM model produces a sharper cosmic web due to its larger velocity perturbation. 

Given this analogy, one can expect that in the cDE models with smaller velocity perturbation the superclusters 
must become less straight. As shown in \citet{baldi12b} and \citet{LB12}, 
the velocity perturbation of the SUGRA003 model is as large as that of the EXP003 case before 
$z\sim 7$ but suddenly changes its tendency, dropping rapidly below that of the $\Lambda$CDM model 
at lower redshifts \citep[see Figure 7 in ][]{LB12}.  Therefore, at present epoch the velocity perturbation in the 
SUGRA003 model is even smaller than that of the $\Lambda$CDM case, which leads to blunt the cosmic 
web and to diminish the degree of the straightness of the supercluster spines. As for the EXP002 and 
EXP003 model with constant coupling, the interplay between the dark sector coupling and the larger 
velocity perturbation determines the decrement in the degree of the supercluster straightness: The 
presence of the dark sector coupling tends to undermine the degree of the supercluster straightness while 
the large velocity perturbation relative to the $\Lambda$CDM case plays a role in straightening the 
superclusters. Our result shown in Figure \ref{fig:mssize} indicates that the former effect should be more 
dominant. 

\begin{figure}
\begin{center}
\plotone{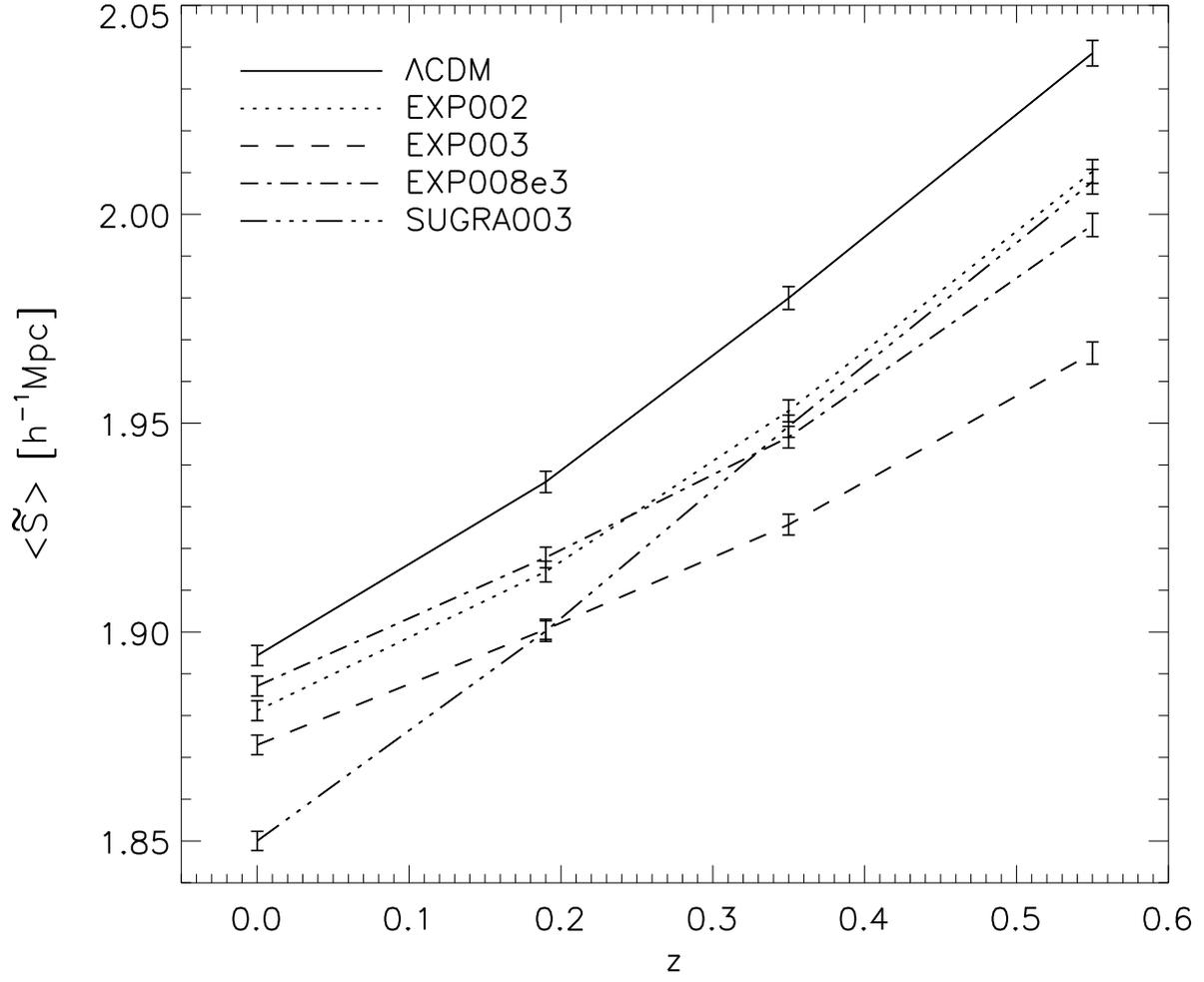}
\caption{Redshift evolution of the mean specific sizes of the supercluster spines for the five models. }
\label{fig:mssize_z}
\end{center}
\end{figure}

To investigate how the supercluster straightness evolves for each model, we repeat the whole process at three 
higher redshifts: $z=0.19,\ 0.35,\ 0.55$. Figure \ref{fig:mssize_z} plots the mean 
specific sizes of the supercluster spines as a function of $z$ for five different models. As can be seen, 
for all of the five cosmological models, the specific sizes of the supercluster spines increase with redshifts. 
It can be well understood by the fact that at higher redshifts the superclusters obtained as FoF groups 
from the mass-limited cluster sample (with the same mass threshold of $10^{13}\,h^{-1}M_{\odot}$) 
correspond to more linear regimes where the clustering of clusters is not so strong. 

As can be also seen in Figure \ref{fig:mssize_z}, at all redshifts, the $\Lambda$CDM case is found to have the 
largest mean specific sizes of the supercluster spines. The strongest evolution in $\langle\tilde{S}\rangle$  is 
found for the SUGRA003 case, while the EXP003 shows the weakest evolution.  It is also interesting to see that 
at higher redshifts the EXP008e3 case differs significantly in $\langle\tilde{S}\rangle$ from the $\Lambda$CDM 
case. As shown in \citet{LB12}, the velocity perturbation in the EXP008e3 model increases exponentially relative 
to the $\Lambda$CDM case at lower redshifts. Therefore, the significant difference in $\langle\tilde{S}\rangle$ 
between the EXP008e3 and the $\Lambda$CDM case at high redshifts must be due to the smaller velocity 
perturbation of the EXP008e3 at higher redshifts than at the present epoch, which is consistent with our interpretation that the 
larger (smaller) velocity perturbation functions for (against) sharpening the cosmic web.

The result shown in \ref{fig:mssize_z} also shows that at higher redshifts the differences in 
$\langle\tilde{S}\rangle$ between the cDE and the $\Lambda$CDM cases become more significant. As 
mentioned in the above, the high-$z$ superclusters are more linear objects and thus they are more vulnerable to 
the long-range fifth force in cDE models. The crucial implication of this result is that the redshift evolution of 
$\langle\tilde{S}\rangle$ must be a powerful complimentary probe of cDE.

\section{SUMMARY AND DISCUSSION}

Although the dependence of the "filamentarity" of the rich superclusters on the initial conditions of the Universe 
was already noted by several authors, \citep[e.g.,][]{dekel-etal84,kolokotronis-etal02,LP06}, the supercluster 
shape distribution was not seriously considered as an efficient cosmological probe mainly because the rich 
filament-like superclusters were regarded  too rare to provide good-number statistics. However, the recently 
available large datasets from the full sky galaxy surveys and the high-resolution simulations have allowed us to 
explore systematically what a new window the noticeable filamentary shapes of the rich supercluster 
can open on the early universe.
 
In this paper we have investigated how the presence of the dark sector coupling in cDE models changes the 
intrinsic clustering pattern of the clusters in the supercluster environments by utilizing the group catalogs from 
the CODECS.  To single out the effect of the dark sector coupling on the clustering of supercluster clusters 
from the nonlinear growth of the large-scale density field, instead of dealing with  the overall filamentary shapes 
of the superclusters, we focus only on the straightness of the main stems (i.e., spines) 
of the minimal spanning trees constructed out of the supercluster clusters. The degree of the straightness 
of a supercluster has been quantified by the specific size of its spine  (spatial extent of the spine per 
member cluster). It has been finally shown that the stronger dark sector coupling makes the superclusters less 
straight and that in the presence of cDE governed by the supergravity potential \citep{BM99} the superclusters 
are least straight.  The difference in the mean specific size of the supercluster spines between the 
$\Lambda$CDM and the viable  cDE models has been found to become more significant  at higher redshifts.

Our results have been physically explained as follows. The attractive fifth force in the cDE models helps the 
large-scale gravitational clustering blunt the cosmic web while the anti-gravitational action of dark energy 
sharpens it.  Since  the supercluster straightness depends on how sharp the cosmic web is, the cDE models 
have less straight superclusters than the $\Lambda$CDM case. The degree of the supercluster straightness 
also depends on the peculiar velocity perturbation of clusters. The large peculiar velocity perturbation of clusters 
functions against blunting the cosmic web, contributing to the degree of the supercluster straightness. 
At higher redshifts, the superclusters for both of the cDE and the $\Lambda$CDM models are more straight 
since they correspond to more linear regimes where the isotropic stress is less dominant. The sharp 
increase of the supercluster straightness with redshifts found in the supergravity model is closely related to 
the bouncing behavior of dark energy equation of state in the supergravity model \citep{baldi12b}. Finally, we 
conclude that the redshift evolution of the supercluster straightness should in principle become a complimentary 
new test of cDE. 

For a practical test of cDE with the evolution of the supercluster straightness, however, 
it will be necessary to deal with the superclusters identified not in real space but in redshift space. 
If the host superclusters are elongated along the line-of-sight directions, the cluster bulk motions along 
the supercluster major axes would cause significant uncertainty on the measurements of the 
supercluster sizes. Furthermore, the effect of the redshift distortion caused by the cluster bulk motion  on the 
supercluster straightness  is likely related to the strength of dark sector coupling since in the cDE models the 
cluster bulk motions must be more active \citep[e.g.,][]{LB12}. It will be definitely important to account for the 
redshift distortion effect on the supercluster straightness and to examine how strongly the effect depends on the 
strength of dark sector coupling. 

The other thing that it will be worth exploring is the robustness of our result against the algorithm to identify 
the filamentary structures in the cosmic web. In the current work, the MST algorithm has been exclusively 
utilized to determine the most prominent filamentary part of the richest section of the cosmic web. But, 
as mentioned in section 2, there exist several more recently developed algorithms which are expected 
to trace more rigorously the linear patterns of the cosmic web on the scales beyond the 
superclusters \citep{skeleton08,mmf10,nexus13}.  If a consistent result be drawn even with a different 
algorithm, it would confirm strongly the usefulness of the supercluster straightness as a probe of coupled dark 
energy. Our future work is in this direction. 
 
\acknowledgments

We thank an anonymous referee for his/her help comments which helped us improve the original manuscript. 
This research was supported by Basic Science Research Program through the National Research Foundation 
of Korea(NRF) funded by the Ministry of Education (NO. 2013004372) and partially by the research grant from 
the National Research Foundation of Korea to the Center for Galaxy Evolution Research  (NO. 2010-0027910).

\clearpage
\clearpage
\clearpage
\clearpage
\clearpage
\clearpage
\clearpage
\clearpage
\clearpage
\begin{deluxetable}{ccc}
\tablewidth{0pt}
\setlength{\tabcolsep}{5mm}
\tablecaption{Numbers of those supercluster spines with three or more nodes and the fraction occupied by 
those spines from which the most massive clusters are pruned away.}
\tablehead{model & $N_{\rm spine}$ & fraction \\
& & $[\%]$} 
\startdata
$\Lambda$CDM  & $26311$ & $2.79$\\
EXP002 & $26678$ & $2.59$\\  
EXP003  & $26636$ & $2.32$ \\
EXP008e3 & $26259$ & $2.51$ \\
SUGRA003 & $28400$ & $2.82$ \\  
\enddata
\label{tab:mmass}
\end{deluxetable}

\end{document}